\def\year{2021}\relax
\documentclass[letterpaper]{article} 
\usepackage{aaai20}  
\usepackage{times}  
\usepackage{helvet} 
\usepackage{courier}  
\usepackage[hyphens]{url}  
\usepackage{graphicx} 
\usepackage{graphicx}  
\usepackage{amsmath,amssymb,amsfonts}
\usepackage{algorithmic}
\usepackage{textcomp}
\usepackage{xcolor}
\usepackage{comment}
\usepackage{todonotes}
\usepackage{booktabs} 
\usepackage{multirow}
\usepackage{comment}

\title{Attention Based Video Summaries of Live Online Zoom Classes}

\author{
Hyowon Lee, Mingming Liu, Hamza Riaz, Navaneethan Rajasekaren,\\ 
\Large \textbf{Michael Scriney, Alan F. Smeaton}\\

 Insight Centre for Data Analytics\\
 Dublin City University,\\ 
 Glasnevin, Dublin 9, Ireland.\\
 alan.smeaton@dcu.ie
}

\begin{document}

\maketitle

\begin{abstract}
This paper describes a system developed to help University students get more from their online lectures, tutorials, laboratory and other live sessions. We do this by logging their attention levels on their laptops during live Zoom sessions and providing them with personalised video summaries of those live sessions. Using facial attention analysis software   we  create personalised video summaries composed of just the parts where a student's attention was below some threshold. We can also factor in other criteria into video summary generation such as parts where the student was not paying attention while others in the class were, and parts of the video that other students have replayed extensively which  a given  student has not. 
Attention and usage based video summaries of live classes are a form of personalised content, they are educational video segments recommended to highlight important parts of live sessions, useful in both topic understanding and in exam preparation.
The system also allows a Professor  to review the aggregated attention levels of those in a class who attended a live session and logged their attention levels. This  allows her to see which parts of the live activity  students were paying most, and least, attention to. 
The Help-Me-Watch system is deployed and in use at our University in a way that protects student's personal data, operating in a GDPR-compliant way.
\end{abstract}

\section{Introduction}

The conventional model of teaching at University level has been changed, possibly forever, as a result of the COVID-19 pandemic.  Prior to this, most Universities and Colleges had  moved to a method of teaching students based on a combination of stand up lectures to large or small classes, smaller group interactive tutorials, laboratory sessions, peer mentoring and others. This was backed up by online resources and access to course materials like notes, presentations, links to online pre-recorded videos, quizzes, and other interactive artefacts.  Online Virtual Learning Environments (VLEs) have emerged as a platform to gather and manage such resources and the use of  systems including Moodle \cite{kumar2011comparative} and Blackboard \cite{heaton2005introducing} had become widespread.  

MOOCs have also played a role in the digitisation of education with Universities putting some or most of their teaching content online for both their own students as well as for a wider population of those interested in learning, either for formal qualification or just for broadening their knowledge \cite{gil2018teachers}. The effect of this has been that the demands of students when they are learning online are different to the on-campus environment and we are still trying to understand what those new requirements are.
Rather than being driven by pedagogical concerns, much of the move to online learning, both the gradual shift over the last decade and the recent stampede as a result of the pandemic, has been done  initially because it was possible and then because it was necessary.  We have not moved online because it was pedagogically correct, and while it brings convenience and reach, as well as economies of scale, we don't know if  the pedagogy should still be  the same  \cite{stephenson2018teaching}.

As reported in \cite{gonzalez2020digital}, who conducted a recent extensive bibliometric analysis of the field, the rapid growth in the use of digital technologies in higher level education is not restricted to just the sciences and engineering disciplines but is right across the field, including the social sciences.
This move to educating students digitally and using digital technologies is partially as a result of the development of technology itself which enables it, and partly attributable to the changing nature of our students.
Today, our students are comfortable with using technology because they've grown up with it and thus they can embrace the use of it as part of their education.  In fact as pointed out in  \cite{cilliers2017challenge}, students as typical representatives of Generation Z are more comfortable with technology than the typical Generation X, who correspond to their Professors.

In early 2020 the higher level education sector, like most others, had to change as a result of the pandemic. We had to pivot from a model of on-campus activities to one where students were online, remote, perhaps had moved back home, and accessing their teaching materials and classes using VLEs and video conferencing  systems like Zoom, Skype or Microsoft Teams.  Because of the rate of spread of COVID-19, this happened almost overnight in most places and with almost zero preparation time, the default was to continue with  scheduled classes and other interactive sessions taking place as online video conferences.

While this helped to get most of us through to the end of that teaching semester, the feedback from students was that many found it difficult to maintain interest and motivation for attending  online lectures over Zoom because they were isolated from others, working alone, and online synchronous lectures were a lacklustre equivalent of face-to-face sessions that did not transfer well \cite{student-feedback}. This illustrated that the pedagogy of online is not the same as on-campus.  Many also had  internet access difficulties.

As a result of this experience, many Universities have since moved to  flipped classrooms or hybrid delivery as a form of blended learning \cite{van2020idea} where students are required to prepare for online classes in advance through pre-recorded video lectures or material to be read. The online live classes are supposedly more engaging as they are more interactive and can focus on problem-based learning activities.

However, while this may improve the material, the other factors remain, namely that students are on Zoom, likely working alone and isolated from their peers, and they are easily demotivated.  That means that the online class experience for them is lessened because there are real world distractions, the level of engagement afforded by video-conference sessions as a whole  is poor and the amount of social interaction with others while at an online lecture, is minimal \cite{martin2018engagement}.

In the work introduced in this paper we use AI techniques to address these shortcomings  using the Help-Me-Watch system which is in use at our University.  This highlights for  students, those parts of the live and interactive sessions they missed because although they were attending, they were distracted or not paying sufficient attention.  We do this by generating a personalised video summary of recommended content from a recording of the live session based on their (lack of) attention during that live session.  The approach presents several interesting challenges  in the way video summaries are generated based on  ambient monitoring of students’ attention levels, the playback usage of different parts of the video recording and the ways which aggregated feedback to the Professor can be generated and presented. 

\section{Student Monitoring}

Prior to the arrival of the COVID-19 pandemic, University education had already taken  steps towards virtualisation with the use of online platforms, namely VLEs, for providing online access to learning materials.  This has helped the teaching and learning process by providing easy access as well as allowing interaction through quizzes, class polls and surveys, computer programming environments, etc.  

However notwithstanding the reservations some  have about its use for example \cite{wired2020}, VLE log access data which records student access to online resources has  been used for 
over a decade in a field known as 
learning analytics.  The applications for this are 
early detection of students who are struggling with their learning \cite{azcona2019detecting} as well as personalising the delivery of educational content \cite{azcona2018personalizing} and even in predicting course outcome in terms of final grades in examinations \cite{waheed2020predicting}. 

While this may be regarded as a form of student monitoring, learning analytics has positive connotations compared to other types of person-monitoring, and there are many examples of it having a positive impact on learning and on outcomes.
In general, similar to that reported in \cite{8026185}, students are not affected in their learning despite knowing that they are being monitored ambiently, just like in other aspects of society we do not change our behaviour when we know our activities are logged when we are online, in city spaces with CCTV, etc.
Yet despite the widespread use of learning analytics, the evidence for the success in its use  in improving student learning remains anecdotal rather than systematic \cite{ifenthaler2019utilising} and it will take time for these benefits to become accepted across the board.

\section{Video Summarisation}

Automatic video summarization is the task of generating a short version of a longer video by selecting the highlights or gist of the original video, thus compacting the storyline into a reduced timespace.  It is not a new topic  as this review article from 2008 shows  \cite{MONEY2008121}. Video summarization approaches depend on the genre of the video being summarised meaning that we will adopt different strategies for summarizing different video types.  For example if we are summarizing a video with a storyline, like a movie which is a thriller, then we may not want to reveal the ending of the story, whereas for an action movie we may wish to include the best of the action shots in the summary \cite{10.1145/1178677.1178709}. 

If we want to generate a movie trailer which does not reveal the storyline but includes the scenes with most suspense, as an incentive for the viewer to want to watch the full movie, then we need semantic understanding of the original video as was done with the movie Morgan where the trailer was generated using the IBM Watson system \cite{smith2017harnessing}. 

Generating summaries of sports videos requires a different approach as we want the most exciting moments in the sporting event to be included in the highlights. Using cricket as an example,  \cite{iet:/content/journals/10.1049/iet-ipr.2018.5589} have used a range of cues for determining the highlights including   excitation level as indicated by the pitch of the commentator's voice. 

Other video genres including CCTV footage, egocentric video, TV news or documentaries, and lecture presentations, will each have their own differing criteria as to what should be included in their summary.

The idea of generating a video summary based on direct feedback from viewers as the video is being watched, has been reported previously. For example, using the facial expressions of viewers,
perception based summaries which identify the most affective scenes in videos, have been  generated \cite{10.1145/1646396.1646435}. This approach was tested on 8 short video clips of various genres and a range of emotions were classified from the facial analysis of viewers, including  neutral, happy, surprised, angry, disgust, fear, and sad.  Using this it was shown that it is possible to generate quite elaborate video summaries without requiring analysis of the original video content.

\section{The Help-Me-Watch System}

\begin{figure*}[!t]
    \centering
    \includegraphics[width=0.85\textwidth]{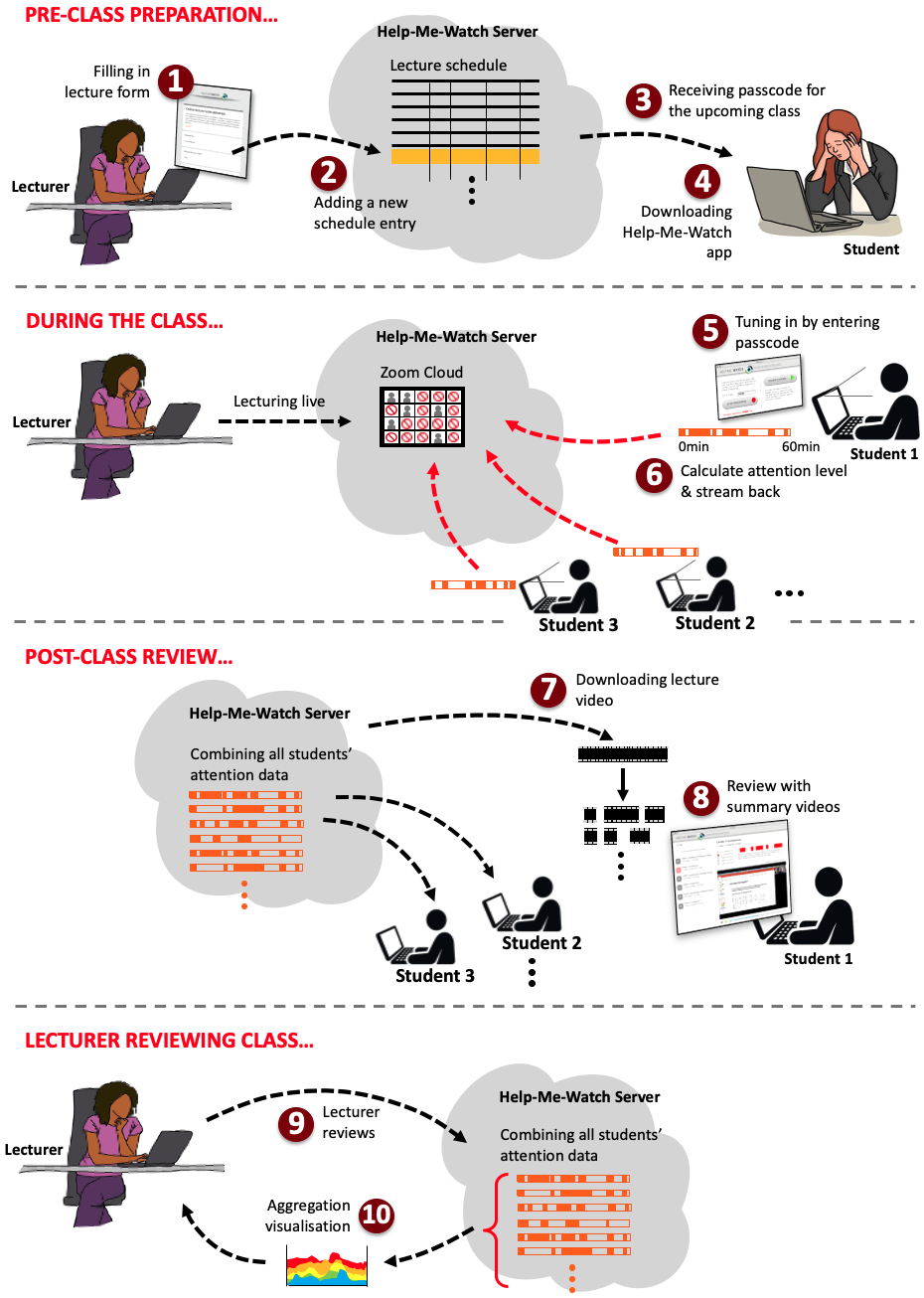}
    \caption{Information flow for the Help-Me-Watch system}
    \label{fig:system}
\end{figure*}

We built and deployed a system which generates personal video summaries  of live online Zoom class content for students, called Help-Me-Watch.
The design and information flow in the  system is shown in Figure~\ref{fig:system}, and it operates in 4 phases.

\begin{itemize}

\item The system begins by inviting a Professor to register their  forthcoming course with the system (1). This generates a unique public passcode for all lectures in the course (2) which is shared with students for them to use for that course (3). Also in advance of the live session, students download and install our Help-Me-Watch application on their laptop (4).  The system also generates a private passcode for the Professor which she keeps private.  

\item {\em During the class}, students and the Professor  connect to Zoom at the appointed time for the live session and students also run the Help-Me-Watch app  on their computer (MAC or Windows) (5)  entering the public passcode.
This downloads the course code and  title and ensures that attention level data from students attending different courses are kept separate. During the live class,  students’ webcams compute their individual attention levels and stream this data back to our server (6) for processing.


\item Some time after the live session, when a student wants to do a {\em post-class review} of the material presented during the live class, the live session which has been recorded in full on the Zoom platform (7) is automatically summarised for that student using their own attention level data, and optionally the attention level data from other students in the class and usage data on which parts of the video other students have played.
Each student is thus able to review their own personalised summary version of the lecture on their own  laptop or mobile device (8). 

\item Also at some point after the class, the {\em Professor can review the class} (9) by entering their private passcode for that course, so students cannot access this facility, and the aggregated and anonymised student attention data for the live session is presented (10) as feedback into what parts of their lecture attracted most, and least, attention from the class. This is presented as a stacked line graph and it is a proxy from the kind of visual body language any presenter would get in front of any audience, except it is retrospective and not live, though live feedback is an option we will pursue in the future.  

\end{itemize}

\begin{figure*}[!t]
    \centering
    \includegraphics[width=0.8\textwidth]{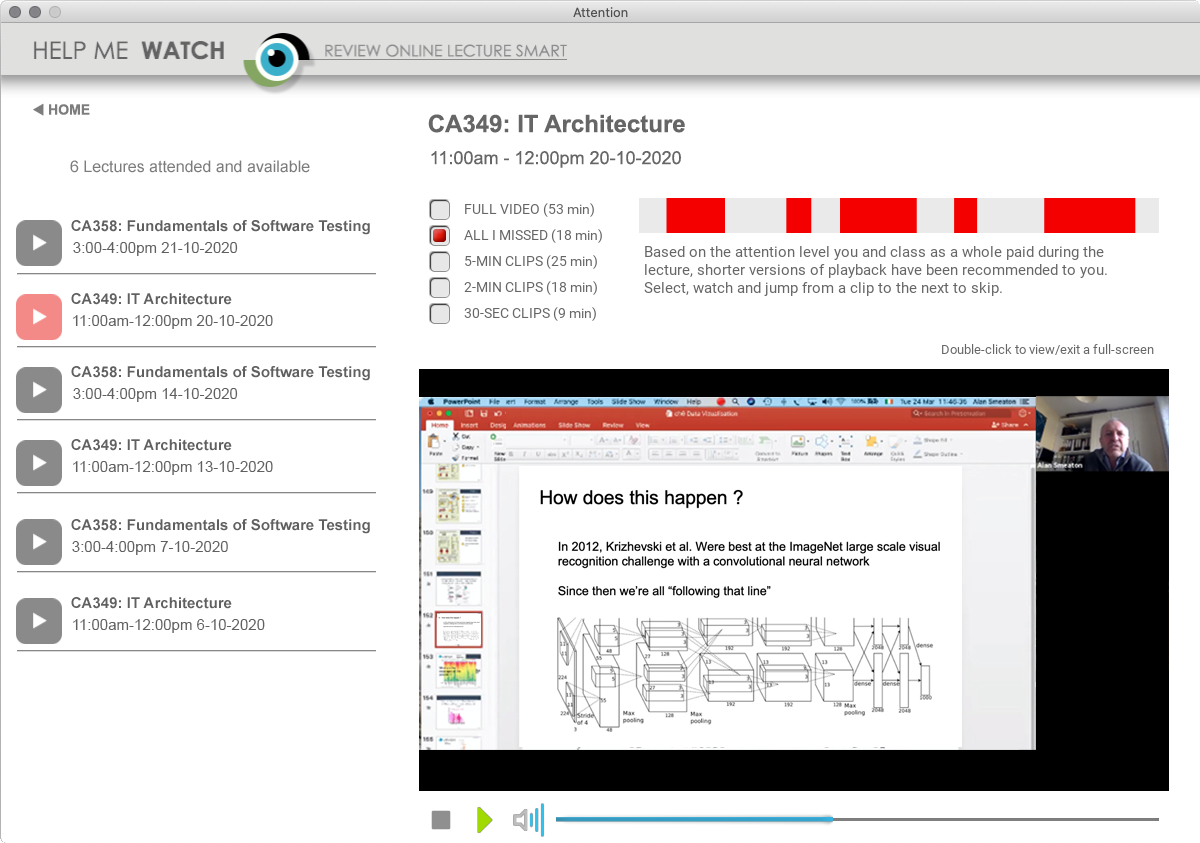}
    \caption{Screengrab of student replaying a video summary of a past lecture}
    \label{fig:student-review-screen}
\end{figure*}

Figure~\ref{fig:student-review-screen} shows a screengrab where a student has used Help-Me-Watch for 6 recorded lectures for courses CA358 and CA349. She has chosen to review the live lecture for {\em CA349 IT Architecture}, a class held on 20th October 2020 between 11am and 12pm.  Figure~\ref{fig:student-review-screen} shows that the student can choose between replaying the full original video (53 minutes duration) or playing just the parts where her attention levels dropped below a threshold (18 minutes duration), or automatically generated video summaries where the individual video segments which are appended together to make the summary are of 5-minutes (25 minutes overall), 2-minutes (18 minutes overall) or 30 seconds (9 minutes overall) duration.  

In the case of Figure~\ref{fig:student-review-screen} the student has chosen to view the 18 minutes of the ``all I missed’’ summary and the actual parts that were missed, or where attention levels dropped below a threshold, are highlighted as red bars on the screen.  The screen also includes an embedded video playback window, with play, pause and stop controls.  The student is about halfway through playing this summary, with the on-screen material containing a description of the convolutional neural network used in the ImageNet challenge in 2012, which is part of the course on IT Architecture.

In addition to using students' attention level data to generate personalised video summaries for each student, the attention level data  is aggregated and summarised with the contributions of students anonymised and can be presented back to the Professor.  For this to happen the Professor uses a second automatically-generated passcode, this time the private passcode, so only the Professor  can access this. 

\begin{figure*}[!t]
    \centering
    \includegraphics[width=0.8\textwidth]{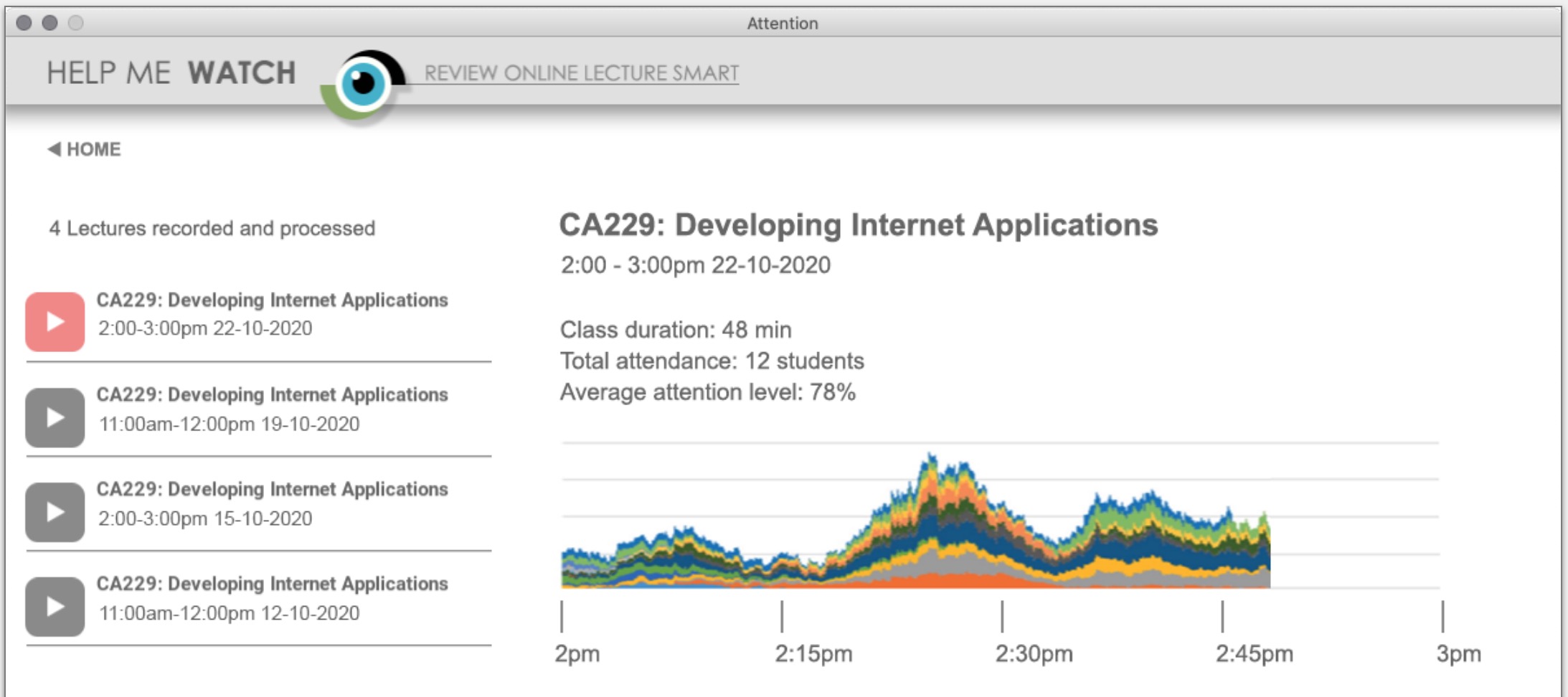}
    \caption{Screengrab for Professor reviewing past lectures}
    \label{fig:lecturer-review-screen}
\end{figure*}

Figure~\ref{fig:lecturer-review-screen} shows a screengrab of the lecture review options for a Professor, indicating she has allowed Help-Me-Watch to be used for 4 of her lectures to date as part of her module CA229: Developing Internet Applications.   She has chosen to highlight the lecture which took place on 22 October 2020 between 2pm and 3pm.  Figure~\ref{fig:lecturer-review-screen} shows that the class was 48 minutes in duration, that 12 students used Help-Me-Watch and the stacked bar chart shows  anonymised aggregated attention levels from those 12 students.  From this we can see that for the first 20 minutes the lecture was of mid-range interest and then got interesting towards the middle part, perhaps because the Professor was giving details of the class assignment.  It then tailed off to about 2:45 before rising again for the remainder of the lecture.  The lesson for this Professor for this particular lecture is that the second half was better than the first  in terms of student attention, and there was something really interesting for these 12 students in the middle.  In a more recent implementation of Help-Me-Watch we have synchronised the stacked line graph of attention levels with a video playback window, similar to what us used in, for example, medial debriefings.  

The Help-Me-Watch system has been built, deployed and is in use at Dublin City University where we are using it to gather usage data and  feedback from Professors and students. In this, the estimation of student attention level which runs on the app downloaded onto students' laptops is based on a real-time eye blink detection algorithm  \cite{Soukupov2016RealTimeEB} that computes the eye aspect ratio (EAR) between height and width of the eye. It does this by estimating the landmark positions around the eye in real time and extracting a single scalar value.  Baseline EAR values across different lectures, both mean and variances, for each student will vary depending on their ethnicity and we use the values specific to a student to determine the thresholds for including clips into their video summaries.  As our dataset and the EAR profiles for individual students grow, we can generate summaries  using not just the overall attention levels across all students attending a class, but also how those attention levels differ from the individual student baselines.

The eye aspect ratio is a  simple algorithm which is accurate and robust and we have tested it in our lab with students of different ethnicities, genders, ages and in different lighting conditions. We have also tested it with and without reading glasses and with and without facial hair. Other more sophisticated real-time methods for attention measurement or even emotion classification could be used but we are satisfied with the robustness of the present implementation. 

However, even though eye gaze as a proxy for attention may be a true reflection of attention in one-to-one conversations between two people either in person or on video conferencing,  when listing to an online presentation a user can be paying attention but not looking at the laptop screen. For example,  when taking notes or perhaps looking at a second, larger monitor on the desk, a student's eye gaze is not fixed at the webcam.  We examine this issue in the next section

\section{Analysis of Usage}

To illustrate Help-Me-Watch in action we  analyse recordings from an online Zoom tutorial session where 9 students from the class used the system to log and upload their attention levels during the 45 minute online tutorial.

\begin{figure*}[!t]
    \centering
    \includegraphics[width=\columnwidth]{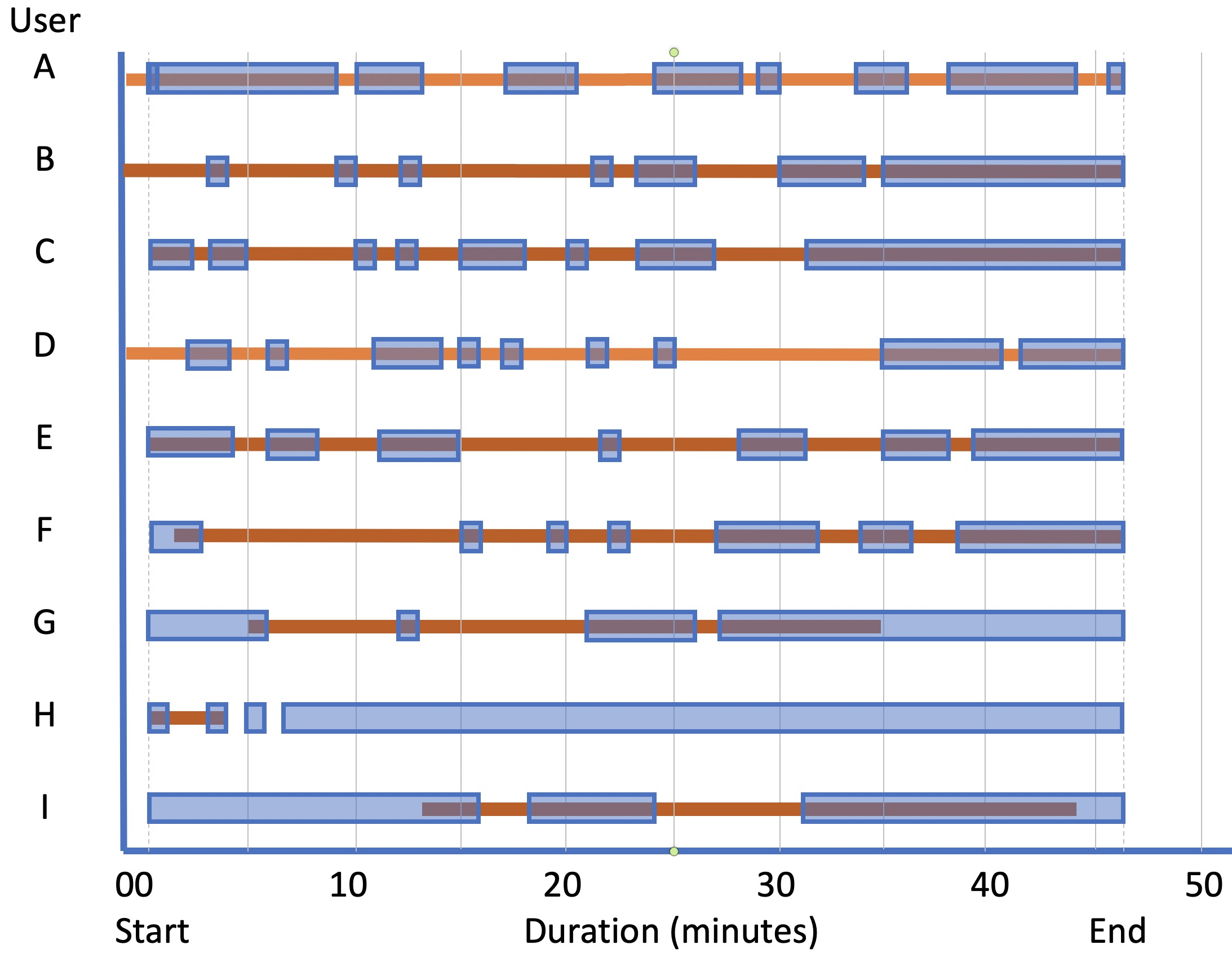}
    \caption{Video segments included in ``all-I-missed'' summaries for 9 students.}
    \label{fig:basline review}
\end{figure*}

Some students had started recording their attention before  the  start  of the video recording by the Professor or continued their recording after the recording ended and we delete these readings before/after the lecture Zoom recording's time boundaries. 
Our baseline video summary approach  used in this example is based on ``all I missed'' and uses  1-minute aggregations of attention levels. Where a one-minute aggregation is in the lower half of observed attention levels for that Zoom session only, that minute is included in the generated summary. Figure~\ref{fig:basline review} and  Table~\ref{tab:baseline} show what these summaries look like for the 9 students.  Note that  where a student's attention level was not  logged in a 1-minute window either because s/he arrived late, left early, or  was not looking at their screen, then the missed part forms part of their video summary.

We will ignore students G, H and I because they attended (or recorded their attention for) a lot less than the full Zoom tutorial duration, the other 6 full attendances generate summaries about which we can say that they are of varied duration, from 16 minutes (D) to 27 minutes (A) for a 47 minute tutorial. They are also  non-contiguous and fragmented, with an average of 8 segments appended together to make the summaries. The segments appearing in the summaries for these 6 students vary from 1 minute (18 such segments) with the longest contiguous segment being 15 minutes (C).

With the fragmented nature and strict cutoffs for stopping and starting segments at 1-minute intervals in our baseline algorithm, these  may be difficult for students to view and comprehend.  We offset that by inserting a 3-second video gap when there is a skip in the video, so as not to disorient students

\begin{table*}[ht]
\centering
\caption{Video segments selected for inclusion in the ``all-I-missed'' baseline summaries for 9 students.}
\label{tab:baseline}
\begin{tabular}{llll}
    \toprule
    Student & Summary segments (minutes) & No. Segments & Duration (min)\\
    \midrule
 A   &  0-9     10-13    17-21   24-28    29-30  34-36  38-44   46-47 & 8 & 27 \\
B & 3-4   9-10   12-13     21-22    23-26   30-34    35-47 &7 & 23\\
C & 0-2  3-5  10-11   13-14   15-18   20-21   23-27    32-47   & 8 &  29 \\
D &2-4  6-7  11-14 15-16  17-18 21-22  24-25    35-41   42-47 & 9 &  16 \\
E &0-4  6-8   11-15    22-23    28-31     35-38      39-47   &7 &  25 \\
F &0-2  15-16  19-20 22-23   27-32   34-36   38-45 &7 &18\\
G &0-6 12-13  21-26 27-47   &4& 32\\
H & 0-1 3-4   5-6   7-47   &4   &38\\
I &0-16  18-24  31-47 &3 &38 \\
    \bottomrule
\end{tabular}
\end{table*}

The baseline summarisation strategy favours including material  below some  computed mean attention level which does not factor in any variance of that attention level for the student or for that Zoom session. As a hypothetical example, we could have a student  with a constant average attention level of 0.3 and  dropping to 0.2 for just the last  minute in which case everything except that last  minute will be above the mean.
We could  address the fragmented nature by varying the threshold so as to reduce the actual number of segments included in the summary but first we will look at the variance in attention for the 9 students.

We regard the series of raw per-second attention levels and the  1-minute attention levels as being a stationary time series  in the sense that the means and variances for each student are constant over time and not subject to some evolving change during a Zoom class.
For such time series,
historical volatility denoted $\sigma$ \cite{somarajan2019modelling,HONG20171}, is a statistical measure 
widely used in economics and finance by analysts and traders in the creation of investing strategies. 
Historical volatility  is the degree of variation over time, usually measured by the standard deviation of logarithmic changes of attention levels.
We computed the raw  per-second volatility of attention levels of the 6 students who attended all of the Zoom tutorial under consideration and we see this in Table~\ref{tab:volatility}

\begin{table}[ht]
\centering
\caption{Video segments selected for inclusion in the ``all-I-missed'' baseline summaries for 9 students.}
\label{tab:volatility}
\begin{tabular}{llll}
    \toprule
    \multirow{2}{*}{Student} & $\sigma$ (per-second & $\sigma$ (1-minute & 1-minute\\
    & attention levels) & attention levels) & volatility\\
    \midrule
 A   &  0.212 & 0.342 & 0.213\\
B & 0.079 & 0.110 & 0.236\\
C & 0.236 & 0.264 & 0.220\\
D &0.085& 0.082 & 0.323\\
E &0.224 & 0.198 & 0.232 \\
F &0.071 & 0.112 & 0.193\\
    \bottomrule
\end{tabular}
\end{table}

This analysis shows  a lot of difference among students in their concentration levels, with $\sigma$ ranging from 0.342 to 0.198,   lower values indicating  consistency in  attention levels. 

In generating a video summary we know there is a tension between one which is choppy and fragmented but includes all the parts missed during the initial online class, compared to a summary which is smoother and with fewer context switches but which is  longer in  duration. If a student is constantly chopping from attending to the online class to focusing on something else or is mind-wandering, then it follows that with a higher volatility measure they will have less focused  attention to the Zoom class. A personal summary will thus have to be either fragmented in nature, or  include large contiguous segments in the summary where the student may have already  paid attention. This is likely to be frustrating to view, somewhat like  re-viewing an old movie or TV episode and realising half-way through that you think you saw this before as you’re remembering parts of it.

A summary generated for a student with low volatility in their attention to an online class is even more unsatisfactory since their differences between attention and non-attention are less pronounced, so it is more difficult to identify which segments to include in the summary.  Thus while our baseline algorithm is a crude first implementation,  this analysis supports the approach of aggregating attention into 1-minute chunks for the purpose of summary generation.

We also consider a student who may be looking at the lecture intently, then looking away to take written rather than typed notes and then look back at the screen, then away to take notes, etc. Here the student's attention levels on a per-second basis will flip or toggle a lot between  high and low attention levels  over a short period of time.  This will be reflected as high volatility in attention  for the 1-minute segment(s) during which this may occur.

\begin{figure*}[ht]
    \centering
    \includegraphics[width=\textwidth]{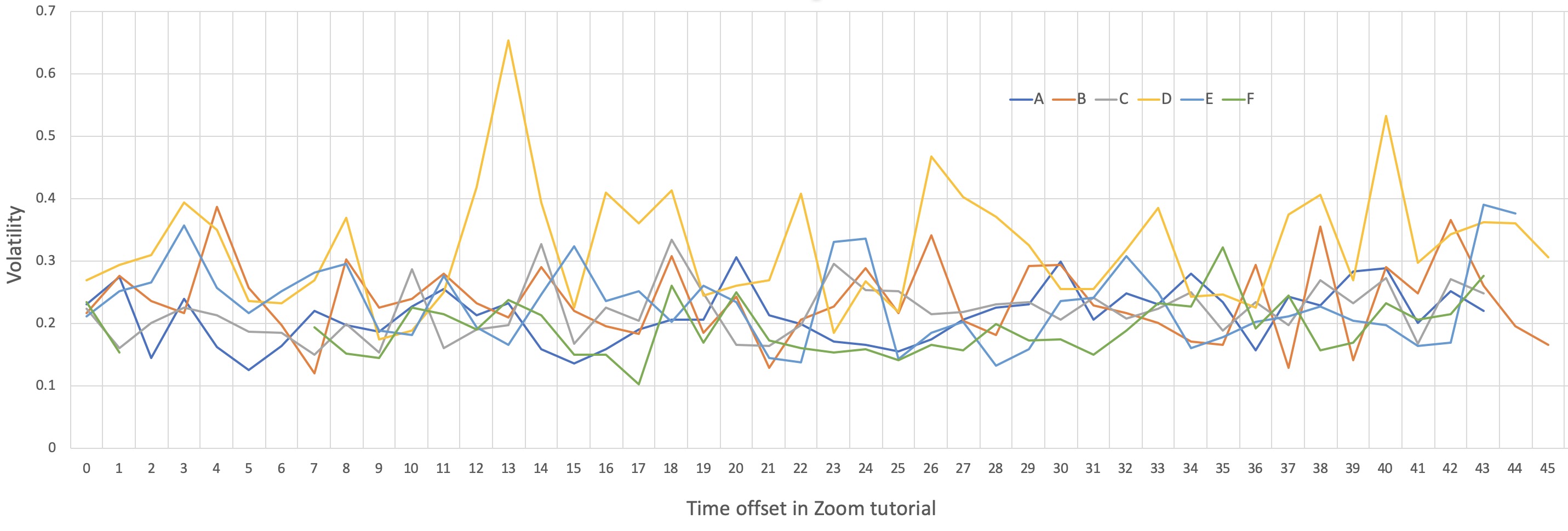}
    \caption{Volatility in attention levels during 1-minute spans}
    \label{fig:volatility}
\end{figure*}

We took all  attention level values for the 6  participants  and  for each participant's 1-minute blocks  we calculated attention volatility for that minute. Figure~\ref{fig:volatility} shows these volatility levels for each participant for each minute during the Zoom session. From this we can see a lot of variability in attention volatility across  participants with participant D (average volatility  0.323) being highest and participant F (average volatility 0.193), these averages  shown in Table~\ref{tab:volatility}.  We  see no correlation among when participants have highly, or low, volatility periods.
The  message from this is that we need to so some observational user experiments to interpret volatility and what is actually causing it in practice but initial interpretation is that this could indicate segments which should be included in summaries.

\section{Future Work}

Some of our plans for future development are engineering improvements while others are more conceptual.  On the engineering side, instead of arbitrarily including 1-minute segments into the  summary, we will introduce some intelligent trimming of  segments incorporated into the summary.  This trimming should be based on pauses in the Professor's dialogue or changes in the slides where they are used.

We do not yet cater for recording attention levels of students attending live online sessions using their smartphones or tablets, just their laptops, though we do support playback and reviewing on smartphones.

Our feedback to the Professor is as a stacked bar chart, colour coded for anonymised students. It is a rich infographic as not only does it show overall class attention (from those who used the Help-Me-Watch app) but  it also shows when students joined and left the session and whether rises in overall class attention are due to the majority of the class or just a small number of  students.  In the case of the feedback to the Professor in Figure~\ref{fig:lecturer-review-screen} we can see that the rise (and subsequent fall) in student attention level around the middle of the lecture is spread almost right across the class so its not just attention from a small subset of students \ldots the whole class was paying attention.

We  plan to include analysis of lecture content, both  audio and visual, to feed back to the Professor what, rather than just where, there were highs and lows in student attention. In the case of  audio, this will follow previous work  on summarization of sports video where the excitation level of the  commentator in terms of voice pitch  is indicative of something exciting happening on the sports broadcast. Similarly we will analyse the visual content  to determine what student attention is  when PowerPoint slides are on-screen for too long and the Professor's face is off-screen, or do animations make a difference to student attention, should slide material be revealed slowly, one bullet point at a time or presented all at once.  Data we are gathering from use of the Help-Me-Watch system will allow personalised feedback to Professors on what features of their presentation style works and what does not work, in terms of grabbing and retaining student engagement.

Finally we would like to offer the Professor the opportunity to flag parts of the content which should be included in everyone's summary, i.e. to indicate the important parts of the lecture which no student should miss.

\section{Conclusions}

In this paper we introduced a system which uses relatively modest AI techniques to generate personalised video summaries of online classes for students to help with class revision and address some of the shortcomings of online learning.  The system called Help-Me-Watch allows educational content to be recommended to students based on their, and in future others students' attention levels during the live classes.  

The system is deployed in a real world setting in our University and actively gathering usage data. This will allow us to do A|B testing to compare the usefulness of different approaches to generating video summaries and a series of planned user studies will give further feedback on the usefulness of the system.

\section{Acknowledgements}
This research was partly supported by Science Foundation Ireland under Grant Number SFI/12/RC/2289\_P2, co-funded by the European Regional Development Fund. The research was also supported by the Google Cloud COVID-19 Credits Program for our cloud based development work.

\bibliography{AAAI-TIPCE-2021-Accepted}
\bibliographystyle{aaai}

\end{document}